\documentclass[aps,pr,twocolumn,floatfix]{revtex4}
\usepackage{graphicx}
\graphicspath{{figures/}}

\usepackage{units,xspace}
\usepackage{amsmath,amssymb}
\usepackage{verbatim}

\newcommand {\bd}[0]{\begin{displaymath}}
\newcommand {\ed}[0]{\end{displaymath}}

\newcommand {\va}{{\sigma_a^2}}
\newcommand {\vb}{{\sigma_b^2}}
\newcommand{\abs}[1]{\left\vert #1 \right\vert}
\newcommand{\avg}[1]{\left< #1 \right>}
\newcommand{\identity}{\mathbf{I}}

\newcommand{\micron}{\ensuremath{\unit{\mu m}}\xspace}

\begin{document}

\title{Optimized holographic optical traps}

\author{Marco Polin}

\author{Kosta Ladavac}

\altaffiliation{Department of Physics, James Franck Institute and
Institute for Biophysical Dynamics, The University of Chicago,
Chicago, IL 60637}
\author{Sang-Hyuk Lee}

\author{Yael Roichman}

\author{David G. Grier}

\affiliation{Dept.\ of Physics and Center for Soft Matter Research,
New York University, New York, NY 10003}

\date{\today}

\begin{abstract}
Holographic optical traps use the forces exerted by computer-generated
holograms to trap, move and otherwise transform mesoscopically textured
materials.
This article introduces methods for optimizing holographic optical
traps' efficiency and accuracy, and an optimal statistical approach for
characterizing their performance.  
This combination makes possible real-time
adaptive optimization.
\end{abstract}


\maketitle

A single laser beam brought to a focus with a strongly
converging lens forms a type of optical trap widely
known as an optical tweezer \cite{ashkin86}.
Multiple beams of light passing simultaneously through
the lens' input pupil yield multiple optical tweezers, each
at a location determined by its beam's angle of incidence and
degree of collimation at the input pupil.
The trap-forming laser beams form an interference pattern as they pass through
the input pupil, whose amplitude and phase corrugations characterize
the downstream trapping pattern.
Imposing the same modulations on a single incident beam at the
input pupil would yield the same pattern of traps.
Such wavefront modification can be performed by a computer-designed diffractive
optical element (DOE) commonly known as a hologram.

Holographic optical trapping (HOT) uses
methods of computer-generated holography (CGH)
to project arbitrary
configurations of optical traps
\cite{dufresne98,reicherter99,liesener00,dufresne01a,curtis02},
with manifold applications in the physical and biological sciences
as well as in industry \cite{grier03}.
This flexible approach to manipulating and transforming
mesoscopic matter has been used to assemble two- and
three-dimensional structures, to sort objects ranging in size from
nanoclusters to living cells, and to create all-optical 
microfluidic pumps and mixers.

This article describes refinements of the HOT technique
that help to optimize the traps' performance, as well as
a statistically optimal analysis for rapidly
characterizing them.
Section~\ref{sec:optics} describes modifications to the basic
HOT optical train that minimize defects due to limitations of
practical implementations.  Section~\ref{sec:algorithm}
discusses a direct search algorithm for HOT DOE computation that
is both faster and more accurate than iterative
refinement algorithms.  
A method for rapidly characterizing each trap in a holographic array
is presented in Section~\ref{sec:characterization}.
The optimal statistical methods on which this characterization technique
is based lends itself naturally to digital video analysis 
of optically trapped spheres and can be exploited for real-time
optimization.
Such adaptive optimization is demonstrated experimentally in
Section~\ref{sec:experiment}.

\section{Improved optical train}
\label{sec:optics}

\begin{figure}[tbhp]
  \centering
  \includegraphics[width=3in]{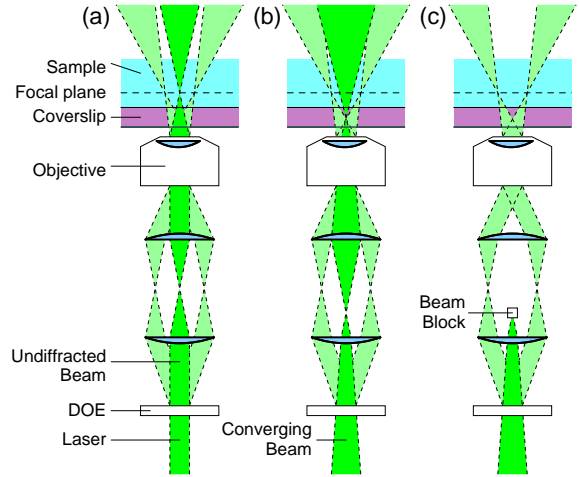}
  \caption{Simplified schematic of a holographic optical tweezer optical train
    before and after modification.  (a) A collimated beam is split into multiple
    beams by the DOE, each of which is shown here as being collimated.  
    The diffracted beams
    pass through the input pupil of an objective lens and are focused into
    optical traps in the objective's focal plane.  The undiffracted portion of
    the beam, shown here with the darkest shading, also focuses into the focal plane.
    (b) The input beam is converging as it passes through the DOE.  The DOE
    collimates the diffracted beams, so that they focus into the focal plane, as in
    (a).  The undiffracted beam comes to a
    focus within the coverslip bounding the sample.
    (c) A beam block can eliminate the undiffracted beam without substantially
    degrading
    the optical traps.
  }
  \label{fig:HOT}
\end{figure}

Figure~\ref{fig:HOT}(a) shows a simplified schematic of a conventional
HOT implementation.  A collimated laser beam
is imprinted with a CGH and thereafter propagates as a superposition
of independent beams, each with individually specified wavefront
characteristics \cite{liesener00,curtis02}.
These beams are relayed to the input aperture of a
high-numerical-aperture lens, typically a microscope objective,
that focuses them into optical traps.
This figure indicates a transmissive DOE, although comparable results are
obtained with reflective DOE's.
The same objective lens used to form the optical traps also can be
used to create images of trapped objects.  The associated illumination
and image-forming optics are omitted from Fig.~\ref{fig:HOT} for
clarity.
Practical holograms only diffract a portion of the incident light
into intended modes and directions.
Some of the incident beam is not diffracted at all, and
the undiffracted portion typically
forms an unwanted trap in the middle of the field of view \cite{korda02}.
This ``central spot'' has been removed in previous studies by
spatially filtering the diffracted beam \cite{korda02,korda02b}.
Practical DOE's also tend to project
spurious ``ghost'' traps into symmetry-dictated positions within
the sample.
Spatially filtering a large number of
ghost traps generally is not practical, particularly in
the case of dynamic holographic optical tweezers whose traps move
freely in three dimensions.
Projecting holographic traps in the off-axis Fresnel geometry, rather than
the Fraunhofer geometry described here, automatically eliminates the
central spot \cite{jesacher04}, but
limits the number of traps that
can be projected, and also does not address the formation of ghost traps.

Figure \ref{fig:HOT}(b) shows one simple improvement to the basic HOT
design that minimizes the central spot's influence and effectively
eliminates ghost traps.
Here, the source laser beam is converging as it
passes through the DOE.  As a result, the undiffracted 
central spot focuses upstream of the objective's normal focal plane.
The degree of collimation of each diffracted beam, and thus
the axial position of the resulting trap, can be
adjusted by incorporating wavefront-shaping
phase functions into the hologram's design \cite{curtis02}, thereby returning
the traps to the focal volume.
This simple expedient allows the central spot to be 
projected into the coverslip bounding a sample, rather than the sample itself,
thereby ensuring that the undiffracted beam
lacks both the intensity and the gradients needed 
to influence a sample's dynamics.

An additional consequence of the traps' displacement relative to the
converging beam's focal point is that the majority of ghost traps are projected
to the far side of this point, and therefore out of the sample volume
altogether.
This is a substantial improvement for processes such as optical
fractionation \cite{ladavac04,pelton04a}, which rely on
a precisely specified optical potential energy landscape.

Even though the undiffracted beam may not create an actual trap in
this modified optical train, it still can exert radiation pressure on
the region of the sample near the center of the field of view.  This
is a particular problem for large arrays of optical traps in that the central
spot can be brighter than the intended traps.
Illuminating the DOE with a diverging beam \cite{jesacher04a}
reduces the undiffracted beam's influence by projecting some
of its light out of the optical train.
In a thick sample, however, this has the deleterious effect of
projecting both the weakened central spot and the undiminished 
ghost traps into the sample.

These problems all can be mitigated by placing a beam block 
as shown in Fig.~\ref{fig:HOT}(c) in the
intermediate focal plane within the relay optics to spatially filter
the undiffracted portion of the beam.
Because the trap-forming beams come to a focus in a different plane,
they are only slightly occluded by the beam block, even if they pass directly along the
optical axis.  The effect of this occlusion is minimal for
conventional optical tweezers and can be compensated by
increasing their relative
brightness.

\section{Iterative and Direct Search Algorithms for HOT Calculation}
\label{sec:algorithm}

Holographic optical tweezers' efficacy is determined
by the quality of the trap-forming DOE, which in turn reflects
the performance of the algorithms used in their
computation.
Previous studies have applied holograms calculated
by simple linear superposition of the
input fields \cite{reicherter99}, with best results being obtained
with random relative phases \cite{liesener00,curtis02},
or with variations \cite{liesener00,dufresne01a,curtis02} on the classic
Gerchberg-Saxton and Adaptive-Additive algorithms \cite{soifer97}.
Despite their general efficacy, these algorithms yield
traps whose relative intensities can differ greatly from their
design values, and typically project an unacceptably large fraction of the input
power into ghost traps.
These problems can become acute for complicated three-dimensional
trapping patterns, particularly when the same hologram also is used
as a mode converter to project multifunctional arrays of optical traps
\cite{liesener00,curtis02}.
This section describes faster and more effective algorithms for HOT DOE calculation
based on direct search and simulated annealing.

The holograms used for holographic optical trapping typically
operate only on the phase of the incident beam, and not its amplitude.
Such phase-only holograms, also known as kinoforms, are far more
efficient than amplitude-modulating holograms, which necessarily divert
light away from the traps.
Quite general trapping patterns can be achieved with kinoforms
because optical tweezers rely for their operation on
intensity gradients but not on phase variations.
The challenge is to find a phase pattern in the
input plane that encodes the desired intensity pattern in the
focal volume.

According to scalar diffraction theory, the complex field
$E(r, \psi)$, in the
focal plane of a lens of focal length $f$
is related to the field, $u(\vec{\rho}) \, \exp(i \varphi(\vec{\rho}))$,
in its input plane by
a Fraunhofer transform,
\begin{equation}
  \label{eq:fraunhofer}
  E(\vec{r}) = \int u(\vec{\rho}) \,
  \exp(i \varphi(\vec{\rho})) \,
  \exp\left( - i \frac{k \vec{r} \cdot \vec{\rho}}{2 f}\right)
  \, d^2\rho,
\end{equation}
where $u(\vec{\rho})$ and $\varphi(\vec{\rho})$ are the real-valued
amplitude and phase, respectively,
of the field at position $\vec{\rho}$ in the
input pupil, and $k = 2 \pi / \lambda$ is the wavenumber of
light of wavelength $\lambda$.

If $u(\vec{\rho})$ is the amplitude profile of the input laser beam,
then $\varphi(\vec{\rho})$ is the
kinoform encoding the pattern.  Most practical DOEs, including those
projected with SLMs, consist of an array $\vec{\rho}_j$ of discrete
phase pixels, each of which can impose any of $P$ possible 
discrete phase shifts
$\varphi_j \in \{0,\dots,\phi_{P-1}\}$.
The field in the focal plane due to such an $N$-pixel DOE is, therefore,
\begin{equation}
  \label{eq:discrete}
  E(\vec{r}) = \sum_{j = 1}^N u_j \, \exp(i \varphi_j) \, T_j(\vec{r}),
\end{equation}
where the transfer matrix describing the light's propagation from
input plane to output plane is
\begin{equation}
  \label{eq:transfer}
  T_j(\vec{r}) = 
  \exp\left(-i \frac{k \vec{r} \cdot \vec{\rho}_j}{2f} \right).
\end{equation}

Unlike more general holograms, the desired field in the output plane of
a holographic optical trapping system consists of $M$ discrete bright
spots located at $\vec{r}_m$:
\begin{align}
  \label{eq:output}
  E(\vec{r}) & = 
  \sum_{m=1}^M E_m(\vec{r}), \quad \text{with}\\
  \label{eq:fieldm}
  E_m(\vec{r}) & = \alpha_m \, \delta(\vec{r} - \vec{r}_m) \, \exp(i \xi_m),
\end{align}
where $\alpha_m$ is the relative amplitude of the $m$-th trap, 
normalized by $\sum_{m=1}^M \abs{\alpha_m}^2 = 1$, and
$\xi_m$ is its (arbitrary) phase.
Here $\delta(\vec{r})$ represents the amplitude profile of the
focused beam of light, which may be treated at
least heuristically as a two-dimensional Dirac delta function.
The design challenge is to solve Eqs.~(\ref{eq:discrete}), (\ref{eq:transfer})
and (\ref{eq:output}) for the set of phase shifts $\varphi_j$ yielding
the desired amplitudes $\alpha_m$ at the correct locations
$\vec{r}_m$, given $u_j$ and $T_j(\vec{r}_m)$.

The Gerchberg-Saxton algorithm and its generalizations, such as the
adaptive-additive algorithm, iteratively solve both the forward
transform described by Eqs.~(\ref{eq:discrete}) and (\ref{eq:transfer}), and
also its inverse, taking care at each step to converge the calculated
amplitudes at the output plane to the design amplitudes and to
replace the back-projected amplitudes, $u_j$, at the input plane with the
laser's actual amplitude profile.
Appropriately updating the calculated input and output amplitudes at
each cycle can cause the DOE phase $\varphi_j$ to converge to an
approximation to the ideal kinoform, with monotonic convergence possible
for some variants \cite{soifer97}.
The forward and inverse transforms mapping the input and output planes to
each other typically are performed by
fast Fourier transform (FFT).
Consequently, the output positions $\vec{r}$ are
discretized in units of
the Nyquist spatial frequency.
The output field, furthermore, is calculated not only at the intended
positions of the traps, but also at the spaces between them.  This is
useful because the process not only maximizes the
fraction of the input light diffracted into the desired locations, but
also minimizes the intensity of stray light elsewhere.
The down side to this is the additional computational cost for computing
fields in otherwise uninteresting places.

FFT-based iterative algorithms have additional drawbacks for computing
three-dimensional arrays of optical tweezers, or mixtures of more general
types of traps.  To see this, we review how a beam-splitting DOE
can be generalized to include wavefront-shaping capabilities.
A diverging or converging beam at the input aperture comes to a focus
and forms a trap downstream or upstream of the focal plane, respectively.
Its wavefront at the input plane is characterized by the 
parabolic phase profile
\begin{equation}
  \label{eq:curvature}
  \varphi_z(\vec{\rho},z) = \frac{k \rho^2 z}{f^2},
\end{equation}
where $z$ is the focal spot's displacement along the optical
axis relative to the lens' focal plane.
This phase profile can be used to move an optical trap relative to
the focal plane even if the input beam is collimated by appropriately
augmenting the transfer matrix:
\begin{equation}
  \label{eq:transferz}
  T^z_j(\vec{r}) = T_j(\vec{r}) \, K^z_j,
\end{equation}
where the displacement kernel is
\begin{equation}
  \label{eq:displacement}
  K^z_j = \exp(i \varphi_z(\vec{\rho_j},z)).
\end{equation}
The result, $T^z_j(\vec{r})$, replaces $T_j(\vec{r})$ 
as the kernel of Eq.~(\ref{eq:discrete}).

Similarly, a conventional TEM$_{00}$ beam can be converted
into a helical mode through
\begin{equation}
  \label{eq:helical}
  \varphi_\ell(\vec{\rho}) = \ell \theta,
\end{equation}
where $\theta$ is the azimuthal angle around the optical axis
and $\ell$ is known as the topological
charge.  Such corkscrew-like beams focus to ring-like optical
traps known as optical vortices, which can exert torques as well
as forces.  The topology-transforming kernel
$K^\ell_j = \exp(i \varphi_\ell(\vec{\rho}_j))$
can be composed with the transfer matrix in the same manner
as the displacement-inducing $K^z_j$.
A variety of analogous phase-based mode transformations have
been described, each with applications to single-beam optical
trapping \cite{grier03}.  All can be implemented with
appropriate transformation kernels.
Moreover, different transformation operations can be applied
to each beam in a holographic trapping pattern independently,
resulting in general three-dimensional configurations of diverse
types of optical traps \cite{curtis02}.

Calculating the phase pattern $\varphi_j$ encoding multifunctional
three-dimensional optical trapping patterns requires only a slight
elaboration.
The primary requirement is to measure the actual intensity projected
by $\varphi_j$ into the $m$-th trap at its focus.
If the associated diffraction-generated
beam has a non-trivial wavefront, then it need
not create a bright spot at its focal point.
On the other hand, if we assume that $\varphi_j$ creates the required type
of beam for the $m$-th trap through a phase modulation described
by the transformation kernel $K_{j,m}$, then
applying the inverse operator,
$K_{j,m}^{-1}$, in Eq.~(\ref{eq:discrete})
would restore the focal spot.
This principle was first applied to creating three dimensional
trap arrays \cite{liesener00} in which separate translation kernels were used
to project each desired optical tweezer back to the focal plane
as an intermediate step in each iterative refinement cycle.
Computing the light projected into each plane of traps in this manner
involves a separate Fourier transform for the entire plane \cite{liesener00}.

Much of this effort is eliminated by computing
the field only at the traps' positions, as
\begin{equation}
  \label{eq:improved}
  E_m(\vec{r}_m) = \sum_{j=1}^N K_{j,m}^{-1} \,
  T_j(\vec{r}_m) \, \exp(i \varphi_j),
\end{equation}
and comparing the resulting amplitudes $\alpha_m = \abs{E_m}$
with the design values \cite{curtis02}.
Unlike the FFT-based approach, this per-trap algorithm does not
directly optimize the field between the traps.
However, it also eliminates the need to account for interplane
propagation.
Moreover, if the values of $\alpha_m$ match the design values,
no light is left over to create ghost traps.

The per-trap calculation suffers from its own shortcomings.
The only adjustable parameters in Eqs.~(\ref{eq:fieldm})
and (\ref{eq:improved}) are the relative phases $\xi_m$ of
the projected traps.  These $M-1$ real-valued parameters must
be adjusted to optimize the choice of discrete-valued phase
shifts, $\varphi_j$, subject to the constraint that the 
amplitude profile $u_j$ matches the input laser's.
With so few free parameters, however, finding an optimal hologram
is not likely.

Equation~(\ref{eq:improved}) suggests an alternative approach
for computing DOE functions for discrete HOT patterns.
The operator $K_{j,m}^{-1} \, T_j(\vec{r}_m)$
describes how light in the mode of the
$m$-th trap propagates from position
$\vec{\rho}_j$ on the DOE to the trap's projected position
$\vec{r}_m$ in the lens' focal plane.
If we were to change the DOE's phase $\varphi_j$ at that point,
then the superposition of rays composing the field at $\vec{r}_m$
would be affected.
If the change improves the overall trapping pattern, then we would be
inclined to retain it, and seek other such improvements.
If, instead, the result were less good, we would restore $\varphi_j$
to its former value and look elsewhere.
This is the basis for direct search algorithms.

The simplest direct search involves selecting a
pixel at random from a trial phase pattern, changing its value to any of
the $P-1$ alternatives, and computing the effect on the projected
field.
This operation can be performed efficiently by calculating only
the changes at the $M$ traps' positions.
The updated trial amplitudes then are compared with their design values
and the proposed change is accepted if the overall
error is reduced.
The process is repeated until the result converges
to the design or the acceptance rate for proposed
changes dwindles.

A successful and efficient direct search
requires an effective assessment of errors.
The standard cost function,
$\chi^2 = \sum_{m=1}^M (I_m - \epsilon I^{(D)}_m)^2$,
assesses the mean-squared deviations of the 
$m$-th trap's projected intensity
$I_m = \abs{\alpha_m}^2$ from its design value $I^{(D)}_m$,
assuming an overall diffraction efficiency of $\epsilon$.
It requires an accurate estimate for $\epsilon$ and
places no emphasis on uniformity in the projected traps' intensities.
An alternative proposed by Meister and Winfield \cite{meister02},
\begin{equation}
  \label{eq:cost}
  C = - \avg{I} + f \sigma,
\end{equation}
avoids both shortcomings.
Here, $\avg{I}$
is the mean intensity at the traps and
\begin{equation}
  \label{eq:sigma}
  \sigma = 
  \sqrt{\frac{1}{M} \, \sum_{m=1}^M(I_m - \gamma I^{(D)}_m)^2}
\end{equation}
measures the deviation from uniform convergence to the design
intensities.
Selecting
\begin{equation}
  \label{eq:gamma}
  \gamma = \frac{\sum_{m=1}^M I_m \, I^{(D)}_m}{
    \sum_{m=1}^M \left(I^{(D)}_m\right)^2}
\end{equation}
minimizes the total error and accounts for non-ideal
diffraction efficiency.
The weighting fraction $f$ sets the relative
importance attached to diffraction efficiency versus uniformity.

In the simplest direct search for an optimal phase
distribution, any candidate change that reduces $C$ is
accepted, and all others are rejected.
In a worst-case implementation, the number of trials required 
for practical convergence should scale as $NP$, the product of the
number of phase pixels and the number of possible phase values.
In practice, this rough estimate is accurate if $P$ and $N$ are
comparatively small and if the starting phase function is either uniform
or purely random.  
Much faster convergence can be obtained by starting from
the phase obtained by superposing the desired beams with random relative
phases and ignoring amplitude modulations.
In this case, convergence typically is obtained within $N$ trials,
even for fairly complex trapping patterns, and thus requires a
computational effort comparable to the initial superposition.

\begin{figure}[b!]
  \centering
  \includegraphics[width=3in]{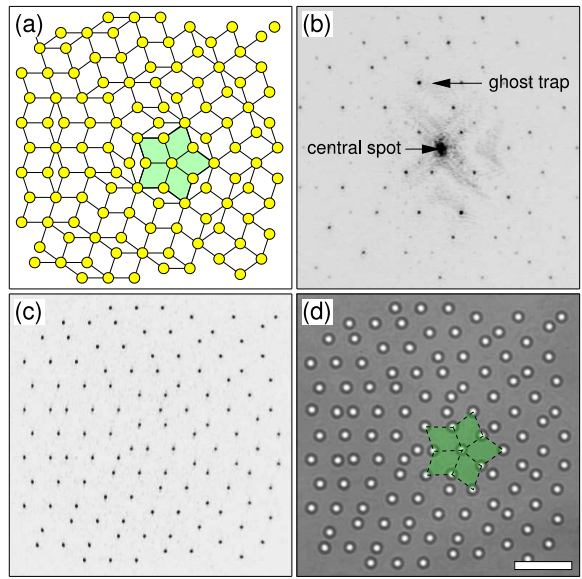}
  \caption{(a) Design for 119 identical optical traps in a
    two-dimensional quasiperiodic array.   (b) Trapping pattern
    projected without optimizations using the adaptive-additive algorithm.
    (c) Trapping pattern projected with optimized optics and
    adaptively corrected direct search algorithm.  (d) Bright-field image
    of colloidal silica spheres 1.53~\micron in diameter dispersed in
    water and organized
    in the optical trap array.  The scale bar indicates 10~\micron}
  \label{fig:quasi}
\end{figure}

As a practical demonstration,
we implement a challenging quasiperiodic array of optical traps.
The traps are focused with a $100\times$ NA 1.4 S-Plan Apo oil immersion
objective lens mounted in a Nikon TE-2000U inverted optical microscope.
The traps are powered by a Coherent Verdi frequency-doubled diode-pumped solid state
laser operating at a wavelength of 532~\unit{nm}.
Computer-generated phase holograms are imprinted on the beam with a Hamamatsu
X8267-16 parallel-aligned nematic liquid crystal spatial light modulator (SLM).
This SLM can impose phase shifts up to $2\pi~\unit{radians}$ at each pixel in
a $768 \times 768$ array.  The face of the SLM is imaged onto the objective's
5~\unit{mm} diameter input pupil using relay optics designed to minimize
aberrations.  The beam is directed into the objective with a dichroic beamsplitter
(Chroma Technologies), which allows images to pass through to a low-noise
charge-coupled device (CCD) camera (NEC TI-324AII).  The video stream is
recorded as uncompressed digital video with a Pioneer 520H 
digital video recorder (DVR)
for processing.

Figure~\ref{fig:quasi}(a) shows the intended planar arrangement of
119 holographic optical traps.
Even after adaptive-additive refinement, the hologram resulting from
simple superposition with random phase fares poorly for this aperiodic pattern.
Figure \ref{fig:quasi}(b) shows the intensity of light
reflected by a front-surface mirror placed in the sample plane.
This image reveals extraneous ghost traps,
an exceptionally bright central spot, and large variability in the
intended
traps' intensities.
Imaging photometry on this and equivalent images produced with different random
relative phases for the beams yields a typical root-mean-square (RMS) variation of
more than 50 percent in the projected traps' brightness.
The image in Fig.~\ref{fig:quasi}(c) was produced using the modified optical train
described in Sec.~\ref{sec:optics} and the direct search algorithm described in 
Sec.~\ref{sec:algorithm}, and suffers from none of these defects.  Both the ghost
traps and the central spot are suppressed, and the apparent relative brightness
variations are smaller than 5 percent, a factor of ten improvement.
Figure~\ref{fig:quasi}(d) shows 119 colloidal silica spheres, $2a = 1.5
\pm 0.3~\micron$ in diameter
(Bangs Labs, lot 5238),
dispersed in water at $T = 27^\circ\unit{C}$ and trapped in the
quasiperiodic array.

\begin{figure*}[!t]
 \centering
 \includegraphics[width=0.75\textwidth]{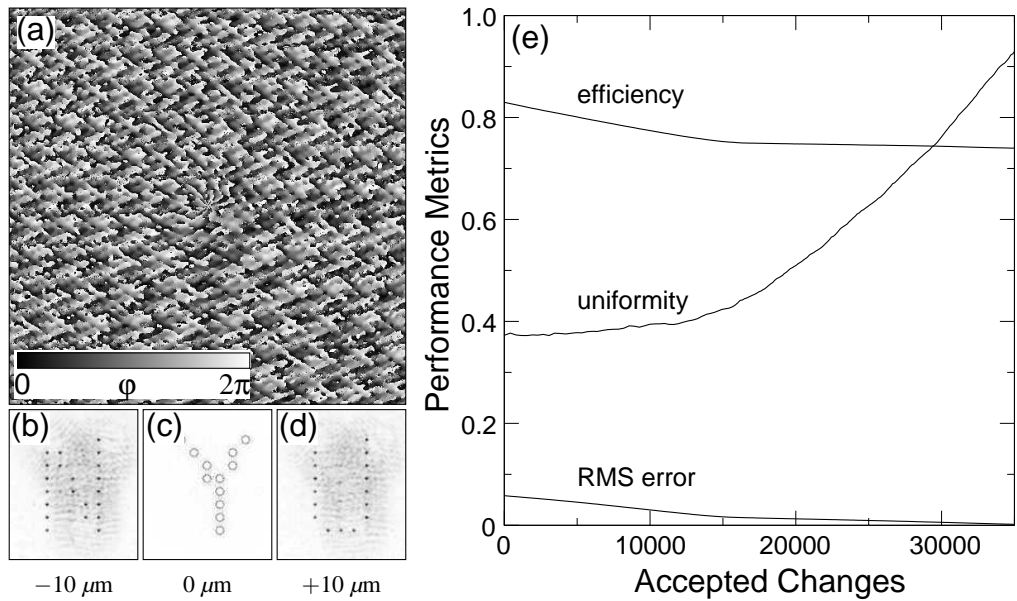}
 \caption{A three-dimensional multifunctional holographic optical trap 
   array created with the
   direct search algorithm.  (a) Refined DOE phase pattern.
   (b), (c) and (d) The projected optical trap array at
   $z = -10~\micron$, $0~\micron$ and $+10~\micron$.
   Traps are spaced by $1.2~\micron$ in the
   plane, and the 12 traps in the middle plane consist of $\ell = 8$
   optical vortices.
  (e) Performance metrics for the hologram in (a)
    as a function of the number of accepted single-pixel changes.
    Data include the DOE's overall diffraction efficiency as defined
    by Eq.~(\ref{eq:efficiency}), the projected pattern's RMS error
    from Eq.~(\ref{eq:error}), and its uniformity, $1 - u$, where
    $u$ is defined in Eq.~(\ref{eq:uniformity}).
  }
  \label{fig:nyumetrics}
\end{figure*}
To place the benefits of the direct search algorithm on a more quantitative
basis, we augment standard figures of
merit with those introduced in Ref.~\cite{meister02}.
In particular, the DOE's theoretical diffraction efficiency is
commonly defined as 
\begin{equation}
  \label{eq:efficiency}
  {\cal Q} = \frac{1}{M} \, \sum_{m=1}^M \frac{I_m}{I^{(D)}_m},
\end{equation}
and its root-mean-square (RMS) error as
\begin{equation}
  \label{eq:error}
  e_\text{rms} = \frac{\sigma}{\max(I_m)}.
\end{equation}
The resulting pattern's departure from uniformity is usefully gauged 
as \cite{meister02}
\begin{equation}
  \label{eq:uniformity}
  u = \frac{\max(I_m/I^{(D)}_m) - \min(I_m/I^{(D)}_m)}{
    \max(I_m/I^{(D)}_m) + \min(I_m/I^{(D)}_m)}.
\end{equation}

Figure~\ref{fig:nyumetrics} shows results for a HOT DOE encoding 51 traps, 
including 12 optical vortices of topological charge $\ell = 8$,
arrayed in three planes relative to the focal plane.  
The excellent results in Fig.~\ref{fig:nyumetrics} were obtained
with a \emph{single} pass of direct-search refinement.
The resulting traps, shown in the bottom three images,
again vary from their planned relative intensities by less than 5
percent.
In this case, the spatially extended vortices were made as bright
as the point-like optical tweezers
by increasing their requested relative brightness by a factor of 15.
This single hologram, therefore, demonstrates
independent control over three-dimensional position, wavefront topology, and
brightness of all the traps.
Performance metrics for the calculation are plotted in 
Fig.~\ref{fig:nyumetrics}(b)
as a function of the number of
accepted single-pixel changes, with an overall acceptance rate of 16 percent.
Direct search refinement achieves greatly improved
fidelity to design over randomly phase superposition
at the cost of a small fraction of the diffraction efficiency
and roughly doubled computation time.
The entire calculation can be completed 
in the refresh interval of a typical liquid crystal spatial
light modulator.

\section{Optimal characterization}
\label{sec:characterization}

At least numerically, direct search algorithms are both faster and better
at calculating trap-forming DOEs than iterative refinement algorithms.
The real test, however, is in the projected traps' ability to trap particles.
A variety of approaches have been developed for gauging the forces exerted
by optical traps.
The earliest involved measuring the hydrodynamic drag required
to dislodge a trapped particle \cite{svoboda93}.
This has several disadvantages, most notably that it identifies
only the marginal escape force in a given direction and not the trap's actual
three-dimensional potential.
Most implementations, furthermore, failed to collect sufficient statistics
to account for thermal fluctuations' role in the escape process, and did not
account adequately for hydrodynamic coupling to bounding surfaces.

Complementary information can be obtained by measuring a particle's thermally
driven motions in the trap's potential well \cite{ghislain94,gittes97,florin98}.
For instance, the measured probability density $P(\vec{r})$
for displacements $\vec{r}$ is related to the trap's potential $V(\vec{r})$
through
the Boltzmann distribution
\begin{equation}
  \label{eq:boltzmann}
  P(\vec{r}) \propto \exp(-\beta V(\vec{r})),
\end{equation}
where $\beta^{-1} = k_B T$ is the thermal energy scale at temperature $T$.
Similarly, the power spectrum of $\vec{r}(t)$ for a
harmonically bound particle is a Lorentzian whose
width is the viscous relaxation time of the particle in the well \cite{ghislain94,bergsorensen04}.

Both of these approaches require amassing enough data to characterize the trapped
particle's least probable displacements, and therefore oversample the trajectories.
Oversampling is acceptable when data from a single optical trap
can be collected rapidly, for example with a quadrant photodiode
\cite{ghislain94,gittes97,florin98,gittes98}.  Tracking multiple particles in holographic optical traps,
however, is most readily
accomplished through digital video microscopy \cite{crocker96}, which yields data
much more slowly.
Analyzing video data with optimal statistics \cite{box76}
offers the benefits of thermal calibration by avoiding oversampling.

An optical trap is accurately modeled as a harmonic
potential energy well \cite{gittes97,florin98,bergsorensen04,gittes98},
\begin{equation}
  \label{eq:harmonic}
  V(\vec{r}) = \frac{1}{2} \, \sum_{i=1}^3 \kappa_i r_i^2,
\end{equation}
with a different characteristic curvature $\kappa_i$ along each axis.
This separable form admits a one-dimensional analysis.
The trajectory of a colloidal particle localized in a viscous fluid
by a harmonic well is described by the one-dimensional Langevin equation \cite{risken89}
\begin{equation}
  \dot{x}(t) = -\frac{x(t)}{\tau} + \xi(t),
\end{equation}
where the autocorrelation time $\tau = \gamma/\kappa$,
is set by the viscous drag coefficient $\gamma$ and the curvature of the well, $\kappa$,
and
where $\xi(t)$ describes Gaussian random thermal forcing with
zero mean, $\avg{\xi(t)} = 0$, and variance
\begin{equation}
  \avg{\xi(t) \xi(s)} = \frac{2 k_B T}{\gamma} \, \delta(t-s).
\end{equation}
If the particle is at position $x_0$ at time $t = 0$, its
trajectory at later times is given by
\begin{equation}
  x(t) = x_0 \, \exp\left(-\frac{t}{\tau}\right) + 
  \int_0^t \xi(s) \, \exp\left(-\frac{t-s}{\tau}\right) \, ds.
  \label{eq:continuous}
\end{equation}
Sampling such a trajectory at discrete times $t_j = j \, \Delta t$, yields
\begin{align}
  x_{j+1} & = x_0 \, \exp\left(-\frac{t_{j+1}}{\tau}\right)
  \nonumber \\ 
  & \quad + \int_0^{t_j} \xi(s) \, \exp\left(-\frac{t_{j+1} - s}{\tau}
  \right) \, ds \nonumber \\
  & \quad + \int_{t_j}^{t_{j+1}} \xi(s) \, \exp\left(-\frac{t_{j+1} - s}{\tau} \right) \, ds \\
  & = \phi \, x_j + a_{j+1} \text{, \quad where \quad}
  \phi = \exp\left(-\frac{\Delta t}{\tau}\right),
  \label{eq:AR}
\end{align}
and where $a_{j+1}$ is a Gaussian random variable with zero mean and variance
\begin{equation}
  \va = \frac{k_B T}{\kappa}\, \left[1 - \exp\left(-\frac{2\Delta t}{\tau}\right)\right].
\end{equation}
Because $\phi < 1$, Eq.~(\ref{eq:AR}) is an example of an autoregressive process
\cite{box76}, which
is readily invertible.
In principle, the particle's trajectory $\{x_j\}$ can be analyzed to extract
$\phi$ and $\va$, and thus the trap's stiffness, $\kappa$,
and the particle's viscous drag coefficient $\gamma$.

In practice, however, the experimentally measured particle positions $y_j$ differ from the
actual positions $x_j$ by random errors $b_j$, which we assume to be taken from
a Gaussian distribution with zero mean and variance $\sigma_b^2$.
The measurement then is described by the coupled equations
\begin{equation}
  x_j = \phi \, x_{j-1} + a_j \text{\quad and \quad} 
  y_j = x_j + b_j,
\end{equation}
where $b_j$ is independent of $a_j$.
We still can estimate $\phi$ and $\va$ from a set of measurements
$\{y_j\}$ by first constructing the joint probability
\begin{align}
  p(\{x_i\}, \{y_i\} \vert \phi, \va, \vb) & = 
  \prod_{j = 2}^N
  \left[\frac{\exp\left(-\frac{a_j^2}{2\va}\right)}{\sqrt{2 \pi
  \va}}\right] \nonumber \\
  & \quad \times \prod_{j = 1}^N \left[\frac{\exp\left(-\frac{b_j^2}{2\vb} \right)}{\sqrt{2 \pi \vb}}\right] \\
  & = \prod_{j=2}^N \left[\frac{\exp\left(-\frac{(x_j-\phi
  x_{j-1})^2}{2\va}\right)}{\sqrt{2 \pi \va}}\right] \nonumber \\
  & \quad \times \prod_{j=1}^N \left[\frac{\exp\left(-\frac{(y_j-x_j)^2}{2\vb}\right)}{\sqrt{2 \pi \vb}}\right].
\end{align}
The probability density for measuring the trajectory $\{y_j\}$, is then
the marginal \cite{box76}
\begin{align}
  p(\{y_j\} \vert \phi,\va,\vb) & = \int p(\{x_j\},\{y_j\} \vert \phi,\va,\vb) \, dx_1 \cdots dx_N
  \\
  & = \frac{(2\pi\va\vb)^{-\frac{N-1}{2}}}{\sqrt{\vb \, \det(A_\phi)}}
  \nonumber \\
  & \quad \times \exp\left(-\frac{1}{2\vb}(\vec{y})^T 
    \left[ \identity -\frac{A_\phi^{-1}}{\vb} \right] \, \vec{y}\right),
\end{align}
where $\vec{y} = (y_1, \dots, y_N)$, $(\vec{y})^T$ is its transpose,
$\identity$ is the $N \times N$ identity matrix, and
\begin{equation}
  A_\phi = \frac{\identity}{\vb} + \frac{M_\phi}{\va},
\end{equation}
with the tridiagonal memory tensor
\begin{equation}
  M_\phi =\left(
    \begin{array}{cccccc}
      \phi^2 & -\phi      & 0          & 0      & \cdots     & 0\\
      -\phi  & 1 + \phi^2 & -\phi      & 0      & \cdots     & \vdots \\
      0      & -\phi      & 1 + \phi^2 & -\phi  & \cdots     & \vdots \\
      0      & 0          & -\phi      & \ddots & \cdots     & \vdots \\
      \vdots & \vdots          & \cdots     & -\phi  & 1 + \phi^2 &-\phi\\
      0      & 0          & \cdots     & 0      & -\phi      & 1
    \end{array}
  \right).
\end{equation}
Calculating the determinant, $\det(A_\phi)$, and inverse, $A_\phi^{-1}$, of
$A_\phi$ is greatly facilitated if we artificially 
impose time translation invariance
by replacing $M_\phi$ with the $(N+1) \times (N+1)$ matrix
\begin{equation}
  \hat{M}_\phi = \left(
    \begin{array}{cccccc}
      1 + \phi^2 & -\phi      & 0          & 0      & \cdots     & -\phi \\
      -\phi      & 1 + \phi^2 & -\phi      & 0      & \cdots     & \vdots \\
      0          & -\phi      & 1 + \phi^2 & -\phi  & \cdots     & \vdots \\
      0          & 0          & -\phi      & \ddots & \cdots     & \vdots \\
      \vdots     & \vdots     & \cdots     & -\phi  & 1 + \phi^2 & -\phi\\
      -\phi      & 0          & \cdots     & 0      & -\phi      & 1 + \phi^2
    \end{array}
  \right).
\end{equation}
Physically, this involves imparting an impulse, $a_{N+1}$, that translates
the particle from its last position, $x_N$, to its first, $x_1$.
Because diffusion in a potential well is a stationary process, the effect 
of this change is inversely proportional to
the number of measurements, $N$.

With this approximation, 
the determinant and inverse of $A_\phi$ are given by
\begin{align}
  \det(A_\phi) & = 
  \prod_{n=1}^{N} \left\{ \frac{1}{\vb} + 
    \frac{1}{\va} \, \left[1 + \phi^2 - 2 \phi \, \cos\left(\frac{2\pi
  n}{N}\right)\right]\right\}  \\
  \text{and} \nonumber \\
  (A_\phi^{-1})_{\alpha \beta} &=
  \frac{1}{N} \, \sum_{n=1}^N 
  \frac{\va \vb \, \exp\left(i \frac{2\pi}{N} n (\alpha-\beta)\right)}{
    \va + \vb \, \left[ 1 + \phi^2 - 2 \phi \, \cos\left(\frac{2\pi n}{N}\right)\right]},
\end{align}
so that the conditional probability for the measured trajectory,
$\{y_j\}$, is
\begin{multline}
  p(\{y_j\} \vert \phi, \va, \vb) = (2\pi)^{-\frac{N}{2}} \\
  \times \prod_{n=1}^{N} \Big\{ \va + 
  \vb \, \left[ 1 + \phi^2 - 2 \phi \, \cos\left(\frac{2\pi n}{N}\right)\right]
  \Big\}^{-\frac{1}{2}} \\
  \times \exp\left(-\frac{1}{2\vb} \, \sum_{n=1}^N y_n^2\right) \\
  \times \exp\left( \frac{1}{2\vb} \, \frac{1}{N} \, 
    \sum_{m=1}^N \frac{\tilde{y}_m^2 \,\va}{
      \va + \vb\, \left[1 + \phi^2 - 2 \phi \, \cos\left(\frac{2\pi m}{N}\right)\right]} \right),
\end{multline}
where $\tilde{y}_m$ is the $m$-th component of the discrete Fourier
transform of $\{y_n\}$.
This can be inverted to obtain the likelihood function for $\phi$, $\va$, and $\vb$:
\begin{multline}
  L(\phi,\va,\vb \vert \{y_i\} ) =
  - \frac{N}{2} \, \ln 2\pi  \\
  - \frac{1}{2\vb} \, \sum_{n=1}^N y_n^2 
 + \frac{\va}{2\vb} \, \frac{1}{N} \,
  \sum_{n = 1}^{N} \frac{\tilde{y}_n^2 \, \va}{
    \va + \vb \, \left[ 1 + \phi^2 - 2 \phi \, \cos\left(\frac{2\pi n}{N}\right)\right]} \\
  - \frac{1}{2} \, \sum_{n=1}^N 
  \ln \left(\va + 
    \vb\left[1 + \phi^2 - 2 \phi \, \cos\left(\frac{2 \pi n}{N}\right) \right] \right)
 .
\end{multline}
Best estimates $(\hat{\phi}, \hat{\va}, \hat{\vb})$ 
for the parameters $(\phi,\va,\vb)$ are solutions of the coupled equations
\begin{equation}
  \frac{\partial L}{\partial \phi} = \frac{\partial L}{\partial \va}
  = \frac{\partial L}{\partial \vb} = 0.
  \label{eq:system}
\end{equation}

\subsection{Case 1: No measurement errors ($\vb = 0$)}
Equations~(\ref{eq:system}) can be solved in closed form if $\vb = 0$.
In this case, 
\begin{equation}
  \hat{\phi}_0  = \frac{c_1}{c_0} \text{\quad and \quad} 
  \hat{\va}_0 = c_0 \, \left[ 1 - \left(\frac{c_1}{c_0}\right)^2\right],
\end{equation}
where 
\begin{equation}
  c_m = \frac{1}{N} \sum_{j=1}^{N} y_j \, y_{(j+m) \bmod N}
\end{equation}
is the barrel autocorrelation of $\{y_j\}$ at lag $m$.
The associated statistical uncertainties are
\begin{equation}
   \Delta \hat{\phi}_0 = \sqrt{\frac{\hat{\va}_0}{N c_0}}
   \text{, \quad and \quad}
   \Delta \hat{\va}_0 = \hat{\va}_0 \, \sqrt{\frac{2}{N}}.
\end{equation}
In the absence of measurement errors, $c_0$ and $c_1$ constitute
\emph{sufficient statistics} for the time series \cite{box76}
and thus
embody all of the information that can be extracted.

\subsection{Case 2: Small measurement errors ($\vb \ll \va$)}
The analysis is less straightforward when $\vb \ne 0$ because
Eqs.~(\ref{eq:system}) no longer are simply separable.
The system of equations can be solved
approximately if $\vb \ll \va$.
In this case, the best estimates for the parameters can be
expressed in terms of the error-free estimates as
\begin{align}
  \hat{\phi} & \approx \hat{\phi}_0 \, \left\{ 1 + \frac{\vb}{\hat{\va}_0} \, \left[
    1 - \hat{\phi}_0^2 + \frac{c_2}{c_0}\right] \right\} \nonumber \\
  \text{and} \nonumber \\
  \hat{\va} & \approx
  \hat{\va}_0 - \frac{\vb}{\hat{\va}_0} \, c_0 \, \left[
    1 - 5 \, \hat{\phi}_0^4 + 
    4 \, \hat{\phi}_0^2 \; \frac{c_2}{c_0} \right],
  \label{eq:firstorder}
\end{align}
to first order in $\vb/\va$,
with statistical uncertainties propagated in the conventional manner.
Expansions to higher order in $\vb/\va$ involve additional 
correlations, and the exact solution involves correlations
at all lags $m$.
If measurement errors are small enough for Eq.~(\ref{eq:firstorder})
to apply, the computational savings relative to other approaches can be substantial,
and the amount of data required to achieve a desired level of accuracy in the physically
relevant quantities, $\kappa$ and $\gamma$, can be reduced
dramatically.

The errors in locating colloidal particles' centroids can be calculated from knowledge of the
images' signal to noise ratio and the optical train's magnification \cite{crocker96}.
Centroid resolutions of 10~\unit{nm} or better can be attained routinely for micrometer-scale
colloidal particles in water using conventional bright-field imaging.
In practice, however, mechanical vibrations, video jitter and other processes may increase
the measurement error.
Quite often, the overall measurement error is most easily assessed by increasing the laser power to the optical
traps to minimize the particles' thermally driven motions.
In this case, $y_j \approx b_j$, and $\vb$ can
be estimated directly.

\subsection{Trap characterization}
\begin{figure*}[t!]
  \centering
  \includegraphics[width=\textwidth]{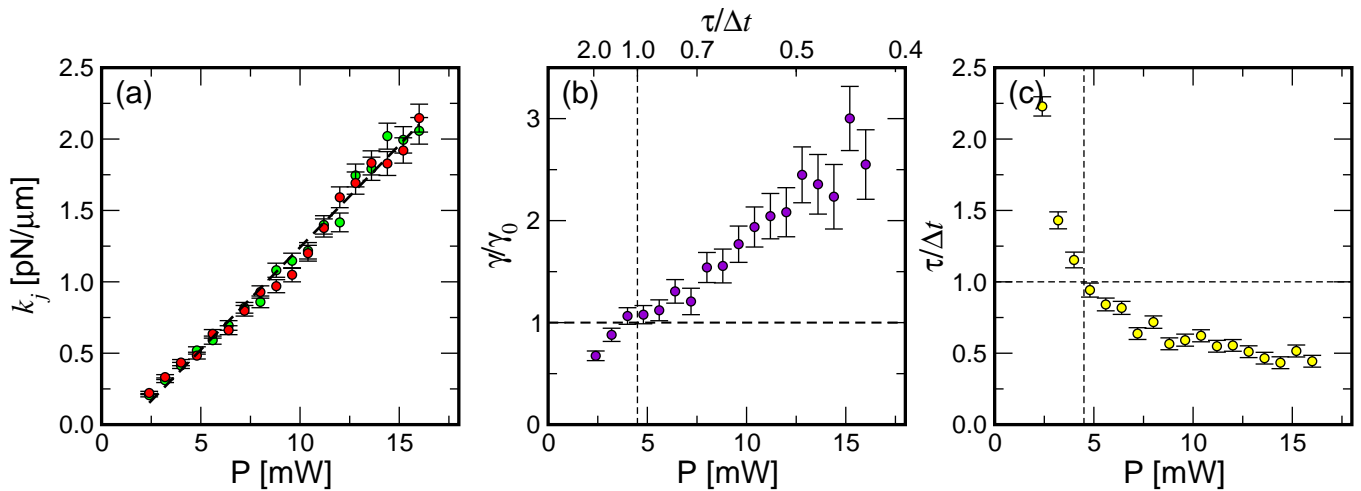}
  \caption{Power dependence of (a) the trap stiffness, (b) 
    the viscous drag coefficient and (c) the viscous relaxation time
    for a 1.53~\micron diameter silica sphere trapped by an optical
    tweezer in water.}
  \label{fig:onesphere}
\end{figure*}
The stiffness and viscous drag
coefficient can be estimated simultaneously as
\begin{equation}
  \frac{\kappa}{k_B T} = \frac{1 - \hat{\phi}^2}{\hat{\va}}
  \text{, \quad and \quad} 
  \frac{\gamma}{k_B T \, \Delta t}  = 
  - \frac{1 - \hat{\phi}^2}{\hat{\va} \, \ln \hat{\phi}}, 
  \label{eq:kcal}
\end{equation}
with error estimates, $\Delta \kappa$ and $\Delta \gamma$, given by
\begin{align}
  \left(\frac{\Delta \kappa}{\kappa} \right)^2 & =
  \left(\frac{\Delta\hat{\va}}{\hat{\va}}\right)^2 +
    \left(\frac{2 \hat{\phi}^2}{1 - \hat{\phi}^2} \right)^2 \,
    \left(\frac{\Delta \hat{\phi}}{\hat{\phi}}\right)^2 \text{ \quad and}\\
    \left(\frac{\Delta \gamma}{\gamma} \right)^2 & =
  \left(\frac{\Delta\hat{\va}}{\hat{\va}}\right)^2 +
    \left(\frac{2 \hat{\phi}^2}{1 - \hat{\phi}^2} +
      \frac{1}{\ln \hat{\phi}} \right)^2 \,
    \left(\frac{\Delta \hat{\phi}}{\hat{\phi}}\right)^2 .
\end{align}
If the measurement interval, $\Delta t$, is much longer than the
viscous relaxation time $\tau = \gamma/\kappa$, then
$\phi$ vanishes and the error in the drag coefficient diverges.  
Conversely, if $\Delta t$
is much smaller than $\tau$, then $\phi$ approaches unity and both
errors diverge.  
Consequently,
this approach does not benefit from excessively fast sampling.
Rather, it relies on accurate particle tracking to minimize $\Delta \hat{\phi}$ and
$\Delta \hat{\va}$.  For trap-particle combinations with viscous relaxation times
exceeding a few milliseconds and typical particle excursions of at least 10~\unit{nm}, 
digital video microscopy provides
the resolution needed to simultaneously characterize multiple optical traps
\cite{crocker96}.

In the event that measurement errors can be ignored ($\vb \ll \va$),
\begin{align}
  \frac{\kappa_0}{k_B T} & = \frac{1}{c_0} \, \left[1 \pm \sqrt{
      \frac{2}{N} \left(1 + \frac{2c_1^2}{c_0^2 - c_1^2}\right)}
      \right]
      \nonumber \\
  \text{and} \nonumber \\
  \frac{\gamma_0}{k_B T \, \Delta t} & = 
  \frac{1}{c_0 \, \ln\left(\frac{c_0}{c_1}\right)} \, \left(
    1 \pm \frac{\Delta \gamma_0}{\gamma_0}\right)
  \label{eq:k0cal}
\end{align}
where
\begin{equation}
  N \, \left(\frac{\Delta\gamma_0}{\gamma_0}\right)^2  =
  2 + \frac{1}{c_0^2 - c_1^2} \,
      \left[ \frac{c_0^2 + 2 c_1^2 \, \ln\left(\frac{c_1}{c_0}\right) - c_1^2}{
          c_1 \ln\left(\frac{c_1}{c_0}\right)} \right]^2.
\end{equation}
These results are not reliable if $c_1 \lesssim \vb$, which arises
when the sampling interval $\Delta t$ is
much longer or much shorter than the viscous relaxation time, $\tau$.
Accurate estimates for $\kappa$ and $\gamma$ still may be obtained
in this case by applying Eq.~(\ref{eq:firstorder}).

As a practical demonstration of this formalism, we analyzed the thermally 
driven motions of a single silica sphere of diameter $1.53~\micron$ 
(Bangs Labs lot number 5328)
dispersed in water and trapped in a
conventional optical tweezer.
With the trajectory resolved to within $6~\unit{nm}$ at 1/30~\unit{sec} intervals, 1 minute
of data suffices to measure both $\kappa$ and $\gamma$ to within 1 percent error using
Eqs.~(\ref{eq:kcal}).
The results plotted in Fig.~\ref{fig:onesphere}(a) indicate trapping efficiencies
of $\kappa_x/P = \kappa_y/P = 142 \pm 3~\unit{pN/\micron W}$.
Unlike $\kappa$, which depends principally on $c_0$, $\gamma$ also
depends on $c_1$, which is accurately measured only for $\tau \gtrsim
1$.
Over the range of laser powers for which this condition holds, we
obtain the expected
$\gamma_x/\gamma_0 = \gamma_y/\gamma_0 = 1.0 \pm 0.1$, as shown in
Fig.~\ref{fig:onesphere}(b).
The viscous relaxation time becomes substantially shorter than our
sampling time for higher powers, so that estimates for $\gamma$ and
for the error in $\gamma$ both become unreliable, as expected.

\section{Adaptive Optimization}
\label{sec:experiment}

Optimal statistical analysis offers insights not only into the
traps' properties, but also into the properties of the trapped
particles
and the surrounding medium.
For example, if a spherical probe particle is immersed in a medium of viscosity $\eta$
far from any bounding surfaces, its hydrodynamic radius $a$ can be assessed from the
measured drag coefficient using the Stokes result $\gamma = 6 \pi \eta
a$.
The viscous drag coefficients, moreover, provide insights into the
particles' coupling to each other and to their environment.
The independently assessed values 
of the traps' stiffnesses then can serve
as a self-calibration in microrheological measurements and in
measurements of colloidal many-body hydrodynamic coupling \cite{polin05a}.
In cases where the traps themselves must be calibrated accurately, knowledge of the
probe particles' differing properties gauged from measurements of $\gamma$ can be
used to distinguish variations in the traps' intrinsic properties from variations due
to differences among the probe particles.

These measurements, moreover, can be performed rapidly enough, even
at conventional video sampling rates, to permit real-time adaptive
optimization of the traps' properties.
Each trap's stiffness is roughly proportional to its
brightness.
So, if the $m$-th trap in an array is intended to receive a fraction
$\abs{\alpha_m}^2$ of the projected light, then 
instrumental deviations 
can be corrected by recalculating the CGH with modified amplitudes:
\begin{equation}
  \alpha_m \rightarrow 
  \alpha_m \, \sqrt{\frac{\sum_{i=1}^N \kappa_i}{\kappa_m}}.
\label{eq:adapt}
\end{equation}
Analogous results can be derived for optimization on the basis of other
performance metrics.
A quasiperiodic pattern similar to that in Fig.~\ref{fig:quasi}(c) was adaptively
optimized for uniform brightness, with a single optimization cycle yielding
better than 12 percent variance from the mean.
Applying Eqs.~(\ref{eq:kcal}) to data from images such as
Fig.~\ref{fig:quasi}(d) allows us to correlate the traps' appearance
to their actual performance.

With each trap powered by 3.4~\unit{mW}, the mean viscous relaxation time
is found to be $\tau/\Delta t = 1.14 \pm 0.11$.
We expect reliable estimates for the viscous drag coefficient under
these conditions, and the result
$\gamma/\gamma_0 = 0.95 \pm 0.10$ with an overall measurement error of
0.01, is consistent with the manufacturer's rated 10 percent
polydispersity in particle radius.
Variations in the measured stiffnesses,
$\avg{\kappa_x} = 0.38 \pm 0.06~\unit{pN/\micron}$ and
$\avg{\kappa_y} = 0.35 \pm 0.10~\unit{pN/\micron}$, 
can be ascribed to a combination of the particles'
polydispersity and the traps' inherent brightness variations.
This demonstrates that adaptive optimization based on the traps' measured
intensities also optimizes their performance in trapping particles.

\section{Summary}
The quality and uniformity of the holographic optical traps projected with
the methods described in the previous sections represent a substantial advance
over previously reported results.
We have demonstrated that optimized and adaptively optimized HOT arrays can be used to
craft highly structured potential energy landscapes with excellent fidelity to design.
These optimized landscapes have potentially wide-ranging applications in
sorting mesoscopic fluid-borne objects through optical fractionation \cite{ladavac04,pelton04a},
in fundamental studies of transport \cite{korda02a,korda02b}, 
dynamics \cite{lee05,lee05a} and phase transitions in macromolecular
systems,
and also in precision holographic manufacturing.

We have benefitted from extensive discussion with Alan Sokal.
This work was supported by the National Science Foundation under
Grant Number DBI-0233971 with additional support from Grant Number
DMR-0451589.  S.L. acknowledges support from the Kessler Family Foundation.


\end{document}